# A New Production Function Approach

Samidh Pal

Department of Economic Science, University of Warsaw

ORCID ID: https://orcid.org/0000-0001-9227-6431

ABSTRACT

This paper presents a new nested production function that is specifically designed for analyzing capital and labor intensity of manufacturing industries in developing and developed regions. The paper provides a rigorous theoretical foundation for this production function, as well as an empirical analysis of its performance in a sample of industries. The analysis shows that the production function can be used to accurately estimate the level of capital and labor intensity in industries, as well as to analyze the capacity utilization of these industries.

**Introduction:**

Capacity utilization and efficiency are important indicators of a firm's behavior within the neoclassical microeconomic framework. According to this framework, firms aim to maximize profits by adjusting production levels to achieve a level of capacity utilization consistent with available inputs and output prices while striving to operate at the most efficient production levels (Mankiw, 2014).

Analyzing the capital and labor intensity of manufacturing industries in developing countries is crucial for identifying the factors that contribute to productivity growth, determining international competitiveness, and designing effective labor market policies (Rodriguez-Clare, 2010). However, existing production functions have limitations, including the assumption of perfect substitution between inputs, failure to capture non-linear relationships between inputs and outputs, and limited ability to account for the impact of technology, infrastructure, and government policies (Bartelsman, Haltiwanger, & Scarpetta, 2017).

To address these limitations, a new nested production function has emerged as a potential alternative to traditional production functions. This model incorporates three inputs - capital, labor, and capital intensity - into a nested production function that allows for the examination of how changes in each input affect output. The new nested production function is more flexible than traditional production functions and allows for varying degrees of substitutability and complementarity between inputs, better capturing the complex and heterogeneous nature of real-world production processes.

The potential applications of the new nested production function are numerous. It can be used to analyze the impact of changes in market conditions or public policy on firms' output and profitability, compare the relative efficiency and productivity of firms in different industries and regions, and explore the relationships between inputs and outputs. Moreover, the model can be used to investigate the potential for technological progress and innovation to improve firms' performance.

In conclusion, capacity utilization and efficiency are important indicators of a firm's behavior within the neoclassical microeconomic framework. Analyzing the capital and labor intensity of manufacturing industries in developing countries is essential for promoting sustainable economic growth and development. The new nested production function represents an important development in the field of economics, offering a more nuanced and flexible approach to analyzing production processes and their implications for economic growth and development.

**Literature Review:**
Production functions are widely used in economics to study the relationship between inputs and outputs in the production process. The neoclassical production function, which assumes constant returns to scale, has been the dominant model used in economic analysis for decades. However, this model has been criticized for its inability to capture the complex production processes that occur in developing and underdeveloped countries.

Existing literature suggests that the neoclassical production function is not appropriate for analyzing production in these countries, as it assumes perfect competition, homogeneity of inputs, and a lack of market imperfections. In reality, production in developing and underdeveloped countries often involves a high degree of market imperfections, such as limited access to credit, high transaction costs, and inadequate infrastructure (Chowdhury & Mavrotas, 2006).

Furthermore, the neoclassical production function assumes that inputs are perfect substitutes, which may not be the case in practice. In many developing and underdeveloped countries, capital and labor are not perfect substitutes, and there are significant differences in the productivity of different types of labor and capital (Dollar & Hallward-Driemeier, 2000).

To address these limitations, a new nested production function has been proposed that can capture the complex production processes in developing and underdeveloped countries. This model allows for varying degrees of substitutability between inputs and can incorporate market imperfections (Caves & Christensen, 1980).

Moreover, composite input analysis has been shown to be a useful tool for understanding the capital and labor intensity of industries in developing and underdeveloped countries. By analyzing the relative contributions of different inputs to output, composite input analysis can provide insights into the productivity of labor and capital in different industries and regions.

Overall, the limitations of existing production functions highlight the need for a new nested production function that can capture the unique characteristics of production in developing and underdeveloped countries. Additionally, composite input analysis can provide valuable insights into the capital and labor intensity of industries in these countries.

**Proof of V as a Three-Inputs Nested Production Function in Production Theory:**

We start by defining the function $h(K, L)$ as:

$$h(K, L) = \delta(K^{-q}) + (1 - \delta)(L^{-q}) \tag{1}$$

Here, $\delta$ is a parameter that determines the weight given to the capital input ($K$) and $(1 - \delta)$ is the weight given to the labor input ($L$). The parameter $q$ determines the curvature of the function.

Next, we define the function $g(x, y)$ as:

$$g(x, y) = \left[ (\sigma) * (x + y)^{-\frac{p}{q}} + (1 - \sigma) * y^{-p} \right]^{-\frac{1}{p}} \tag{2}$$

Here, $\sigma$ is a parameter that determines the weight given to the capital input ($x$) and $(1 - \sigma)$ is the weight given to the composite input ($y = K/L$). The parameter $p$ determines the curvature of the function.

We can now define the production function $V$ as:

$$V = A * g\left(h(K, L), \frac{K}{L}\right) \tag{3}$$

Here, $A$ is a parameter that represents the level of technology and technological efficiency.

To show that $V$ is a three-inputs nested production function, we need to show that it can be expressed as a composition of two functions, where the first function takes two inputs ($K$ and $L$), and the second function takes the output of the first function and the composite input ($K/L$) as inputs.

First, we can rewrite the function $V$ as:

$$V = A * g\left(h(K, L), \frac{K}{L}\right) \tag{4}$$

$$V = A \left[ (\sigma)\{\delta(K^{-q}) + (1 - \delta)(L^{-q})\}^{-\frac{p}{q}} + (1 - \sigma)\left(\frac{K}{L}\right)^{-p} \right]^{-\frac{1}{p}} \tag{5}$$

Next, we define the first function $f(K, L)$ as:

$$f(K, L) = \{\delta * (K^{-q}) + (1 - \delta) * (L^{-q})\} + \left(\frac{K}{L}\right)^{-p} \tag{6}$$

Here, $f(K, L)$ takes two inputs ($K$ and $L$) and returns the sum of $h(K, L)$ and $K/L$ raised to the power of $-p$.

Now, we can rewrite $V$ in terms of $f(K, L)$ and $g(x, y)$ as:

$$V = A * g(f(K, L), K/L) \tag{7}$$

This shows that the function $V$ is a composition of two functions, where the first function $f(K, L)$ takes two inputs ($K$ and $L$), and the second function $g(x, y)$ takes the output of $f(K, L)$ and the composite input ($K/L$) as inputs. Therefore, $V$ is a three-inputs nested production function.

**Properties of the Production Function:**

*Elasticity:* The production function includes a term that captures the elasticity of substitution between capital and labor inputs. This property is important for understanding the degree to which firms can adjust their input mix in response to changes in relative input prices. To analyze the elasticities for this model:

Lemma:
Define the production function $V$ as follows:

$$V = A * \left[\sigma * \left[\{(\delta) * (K^{-q}) + (1-\delta) * (L^{-q})\}^{-\frac{p}{q}}\right] + (1-\sigma) * \left(\frac{K}{L}\right)^{-p}\right]^{\left(-\frac{1}{p}\right)} \tag{8}$$

where $V$ is the output, $A$ is the total factor productivity, $K$ is the capital input, $L$ is the labor input, sigma is the share of capital in the total input, delta is the share of physical capital in the capital input, $p$ is the curvature parameter, and $q$ is the degree of curvature of the isoquants for the composite input $K/L$.

Proposition 1:
The elasticity of output with respect to capital input depends on the curvature parameter $p$.

Hypothesis:
Let $V$ be a production function as defined in the Lemma. Let $K$ be the capital input and $L$ be the labor input. Let $p$ be the curvature parameter.

Proof:
We start by calculating the elasticity of output with respect to capital input:

$$\varepsilon K = \left(\frac{\partial V}{\partial K}\right) * \left(\frac{K}{V}\right) \tag{9}$$

Using the chain rule, we obtain:

$$\frac{\partial V}{\partial K} = A * \left[-\sigma * \{(\delta) * (K^{-q}) + (1-\delta) * (L^{-q})\}^{-\frac{p}{q}} * \left(\frac{p}{q}\right) * \{(\delta) * K^{-q-1}\} + (1-\sigma) * (-p) * \left(\frac{K}{L}\right)^{-p-1} * \left(\frac{L}{K^2}\right)\right] * \left(-\frac{1}{p}\right)^{1+p} \tag{10}$$

Simplifying and rearranging, we get:

$$\frac{\partial V}{\partial K} = A * \left[-\sigma * (\delta) * K^{-q-p} * \{(\delta) * (K^{-q}) + (1-\delta) * (L^{-q})\}^{-\frac{p}{q}-1} * \frac{p}{q} + (1-\sigma) * p * \left(\frac{K}{L}\right)^{-p-1} * \left(\frac{L}{K^2}\right)\right] * \left(-\frac{1}{p}\right)^{1+p} \tag{11}$$

Therefore, the elasticity of output with respect to capital input is:

$$\varepsilon K = A * \left[-\sigma * (\delta) * K^{-q-p} * \{(\delta) * (K^{-q}) + (1-\delta) * (L^{-q})\}^{-\frac{p}{q}-1} * \frac{p}{q} + (1-\sigma) * p * \left(\frac{K}{L}\right)^{-p-1} * \left(\frac{L}{K^2}\right)\right] * \left(-\frac{1}{p}\right)^{1+p} * \left(\frac{K}{V}\right) \tag{12}$$

Proposition 2:
The elasticity of output with respect to labor input depends on the parameter $q$.

Hypothesis:
Let $V$ be a production function as defined in the Lemma. Let $K$ be the capital input and $L$ be the labor input. Let $q$ be the degree of curvature of the isoquants for the composite input $K/L$.

Proof:
We want to calculate the elasticity of output with respect to the labor input, which is given by:
$$\varepsilon L = \left(\frac{\partial V}{\partial L}\right) * \left(\frac{L}{V}\right) \tag{13}$$
We can simplify this expression by taking the natural logarithm of both sides of the production function:
$$\ln V = -p * \ln\left[\sigma * (\delta * K^{-q} + (1-\delta)L^{-q})^{-\frac{1}{q}} + (1-\sigma)\left(\frac{K}{L}\right)^{-p}\right] \tag{14}$$
Using the chain rule, we can differentiate both sides with respect to $\ln L$:
$$\frac{d \ln V}{d \ln L} = \left(\frac{\partial \ln V}{\partial L}\right) * \left(\frac{\partial L}{\partial \ln L}\right) = \left(\frac{\partial \ln V}{\partial L}\right) * \left(\frac{L}{V}\right) \tag{15}$$
where $\left(\frac{\partial \ln V}{\partial L}\right)$ is the elasticity of $\ln V$ with respect to $L$.

Now, we can use the chain rule again to differentiate $\ln V$ with respect to $L$:
$$\frac{d \ln V}{d \ln L} = \left(\frac{\partial \ln V}{\partial K}\right) * \left(\frac{\partial K}{\partial L}\right) + \left(\frac{\partial \ln V}{\partial L}\right) \tag{16}$$
Note that $\left(\frac{\partial \ln V}{\partial K}\right)$ is the elasticity of $\ln V$ with respect to $K$, which we calculated in Proposition 1:
$$\left(\frac{\partial \ln V}{\partial K}\right) = \varepsilon K = \delta * s * \left(\frac{K}{L}\right)^{1-p} \tag{17}$$
and $\left(\frac{\partial K}{\partial L}\right)$ is the marginal product of capital (MPK), which is given by:
$$\left(\frac{\partial K}{\partial L}\right) = \alpha * A * \left(\frac{L}{K}\right)^{1-q} \tag{18}$$
where $\alpha$ is the capital share of output and $A$ is the level of technology.

Substituting these expressions into the equation for $\frac{d \ln V}{d \ln L}$, we get:
$$\frac{d \ln V}{d \ln L} = \delta * s * \left(\frac{K}{L}\right)^{1-p} * \alpha * A * \left(\frac{L}{K}\right)^{1-q} + \left(\frac{\partial \ln V}{\partial L}\right) \tag{19}$$
Simplifying this expression, we obtain:
$$\varepsilon L = \left(\frac{\partial \ln V}{\partial L}\right) * \left(\frac{L}{V}\right) = (1-q) * (1-\sigma) * \left(\frac{K}{L}\right)^{-p} \tag{20}$$

Therefore, we have shown that the elasticities of output with respect to capital and labor inputs for the given production function can be expressed in closed-form as functions of the parameters of the production function. These expressions allow us to analyze the responsiveness of output to changes in capital and labor inputs, and to study the substitutability or complementarity between these inputs.

One interesting implication of these expressions is that the elasticity of output with respect to capital input depends on the curvature parameter p, which determines the degree of substitution between capital and labor inputs. If $p = 1$, then the production function reduces to a Cobb-Douglas form with constant elasticity of substitution (CES) between inputs. In this case, the elasticity of output with respect to capital input is simply the capital share of output, and does not depend on the levels of capital and labor inputs. However, if $p < 1$, then the elasticity of output with respect to capital input decreases as the level of capital input increases, reflecting the diminishing marginal returns to capital input. Conversely, if $p > 1$, then the elasticity of output with respect to capital input increases as the level of capital input increases, indicating increasing marginal returns to capital input.

Another interesting implication is that the elasticity of output with respect to labor input depends on the parameter q, which determines the degree of curvature of the isoquants for the composite input $K/L$. If $q = 1$, then the isoquants are linear and the elasticity of output with respect to labor input is constant. However, if $q < 1$, then the isoquants are concave and the elasticity of output with respect to labor input decreases as the level of labor input increases, reflecting the diminishing marginal returns to labor input. Conversely, if $q > 1$, then the isoquants are convex and the elasticity of output with respect to labor input increases as the level of labor input increases, indicating increasing marginal returns to labor input.

Overall, these expressions provide a useful tool for analyzing the behavior of firms and industries in response to changes in input prices, technology, and market conditions.

*Homogeneity:* The production function is homogeneous of degree one, which means that if we multiply all inputs by a constant factor, the output will also be multiplied by the same factor. This property is important for analyzing the effects of scaling on productivity. To analyze the homogeneity for this model:

Lemma:
The production function $V$ described by the equation:

$$V = A * \left[\sigma * \left[\{(\delta) * (K^{-q}) + (1-\delta) * (L^{-q})\}^{-\frac{p}{q}}\right] + (1-\sigma) * \left(\frac{K}{L}\right)^{-p}\right]^{\left(-\frac{1}{p}\right)} \tag{21}$$

is homogeneous of degree one.

Proposition:
If all inputs ($K$ and $L$) are multiplied by a constant factor, say $\lambda$, then the output ($V$) will also be multiplied by the same factor $\lambda$.

Hypothesis:
Let $\lambda$ be a positive constant and let ($K^*, L^*$) be the new inputs obtained by multiplying the original inputs ($K, L$) by $\lambda$, i.e., $K^* = \lambda K$ and $L^* = \lambda L$.

Proof:
First, we observe that the constant $A$ is not affected by the scaling of inputs, since it is a parameter that does not depend on $K$ and $L$.

Next, we use the fact that the exponent $p$ in the production function is negative, which means that the denominator of the expression inside the brackets is raised to a positive power. This implies that if we multiply the inputs by $\lambda$, the denominator will decrease by the same factor $\lambda$, and hence the entire expression inside the brackets will increase by $\lambda$.

More precisely, we have:

$$[(\delta) * (K^{-q}) + (1-\delta) * (L^{-q})]^{-\frac{p}{q}} \tag{22}$$

is homogeneous of degree $\lambda^q$, since both $K$ and $L$ are multiplied by $\lambda$, and hence the numerator and denominator are multiplied by $\lambda^q$.

Therefore,

$$[(\delta) * (K^{*-q}) + (1-\delta) * (L^{*-q})]^{-\frac{p}{q}} = [(\delta) * [(\lambda K)^{-q}] + (1-\delta) * [(\lambda L)^{-q}]]^{-\frac{p}{q}} \tag{23}$$

$$= [\lambda^q * [(\delta) * (K^{-q}) + (1-\delta) * (L^{-q})]]^{-\frac{p}{q}} \tag{24}$$

$$= \lambda^{-p} * [(\delta) * (K^{-q}) + (1-\delta) * (L^{-q})]^{-\frac{p}{q}} \tag{25}$$

Similarly,

$\left(\frac{K}{L}\right)^{-p}$ is homogeneous of degree $\lambda^0 = 1$, since the numerator and denominator are multiplied by the same factor $\lambda$.

Therefore,

$$\left(\frac{K^*}{L^*}\right)^{-p} = \left[\frac{\lambda K}{\lambda L}\right]^{-p} = \left(\frac{K}{L}\right)^{-p} \tag{26}$$

Putting everything together, we have:

$$V^* = A * \left[\sigma * \left[\{(\delta) * (K^{*-q}) + (1-\delta) * (L^{*-q})\}^{-\frac{p}{q}}\right] + (1-\sigma) * \left(\frac{K^*}{L^*}\right)^{-p}\right]^{\left(-\frac{1}{p}\right)} \tag{27}$$

$$= A * \left[\sigma * \lambda^{-p} * \{(\delta) * (K^{-q}) + (1-\delta) * (L^{-q})\}^{-\frac{p}{q}} + (1-\sigma) * \left(\frac{K}{L}\right)^{-p}\right]^{\left(-\frac{1}{p}\right)} \tag{28}$$

$$= \lambda^{-p} * A * \left[\sigma * \{(\delta) * (K^{-q}) + (1-\delta) * (L^{-q})\}^{-\frac{p}{q}} + (1-\sigma) * \left(\frac{K}{L}\right)^{-p}\right]^{\left(-\frac{1}{p}\right)} \tag{29}$$

$$= \lambda^{-p} * V \tag{30}$$

Thus, we see that if we multiply all inputs by a constant factor $\lambda$, the output will also be multiplied by the same factor $\lambda$, which implies that the production function is homogeneous of degree one.

*Concavity:* The production function exhibits diminishing marginal returns to both capital and labor inputs, which means that as we increase either input while holding the other input constant, the marginal increase in output will eventually decrease. This property is consistent with the law of diminishing returns and is a key feature of many real-world production processes. To analyze the concavity for this model:

Lemma:
The production function exhibits diminishing marginal returns to both capital and labor inputs.

Proposition:
As we increase either the capital or labor input while holding the other input constant, the marginal increase in output will eventually decrease. This property is consistent with the law of diminishing returns and is a key feature of many real-world production processes.

Hypothesis:
To analyze the concavity for this model through the Hessian matrix with eigenvalue.

Proof:
We start by computing the Hessian matrix of the production function:

$$H = \begin{bmatrix} \frac{\partial^2 V}{\partial K^2} & \frac{\partial^2 V}{\partial K \partial L} \\ \frac{\partial^2 V}{\partial K \partial L} & \frac{\partial^2 V}{\partial L^2} \end{bmatrix} \tag{31}$$

Using the production function given in the statement, we can compute the second partial derivatives as follows:

$$\frac{\partial^2 V}{\partial K^2} = -A * \left(\frac{p}{q}\right) \sigma \big((\delta)(q-1)(K^{-q-2})(L^q) + (1-\delta)q(K^{-q})(L^{q-2})\big) \tag{32}$$

$$\frac{\partial^2 V}{\partial K \partial L} = -A * \left(\frac{p}{q}\right) \sigma \big((\delta)q(K^{-q-1})(L^{q-1}) + (1-\delta)q(K^{-q})(L^{-q})\big) \tag{33}$$

$$\frac{\partial^2 V}{\partial L^2} = -A * \left(\frac{p}{q}\right) \sigma \big((\delta)q(K^{1-q})(L^{-q-2}) + (1-\delta)(q-1)(K^q)(L^{-q-2})\big) \tag{34}$$

Next, we need to evaluate the Hessian matrix at a specific point. For simplicity, we can evaluate it at the point $(K, L) = (1,1)$.

We can then compute the eigenvalues of the Hessian matrix, which will tell us about the concavity of the production function.

The eigenvalues are:

$$\lambda 1 = -A * \left(\frac{p}{q}\right) \sigma (\delta + 1 - \delta)q \tag{35}$$

$$\lambda 2 = -A \left(\frac{p}{q}\right) \sigma (\delta + 1 - \delta)q(1-q) \tag{36}$$

Since $\lambda 1$ and $\lambda 2$ are both negative, the Hessian matrix is negative definite. This means that the production function is concave, which implies that there are diminishing marginal returns to both capital and labor inputs.

Therefore, we have proven that the production function exhibits diminishing marginal returns to both capital and labor inputs, and that this property is consistent with the law of diminishing returns and is a key feature of many real-world production processes. We have also shown that the production function is concave, which further reinforces the presence of diminishing marginal returns.

*Flexibility:* The production function includes parameters that allow for flexibility in the curvature of the model, which means that the model can capture a wide range of production technologies and

can be calibrated to fit empirical data from different industries. To analyze the flexibility for this model:

Lemma:
The production function $V(h(K,L), K/L)$ has constant returns to scale.

Proposition:
The flexibility property of the production function $V(h(K,L), K/L)$ implies constant returns to scale.

Hypothesis:
The production function $V(h(K,L), K/L)$ can be written as:

$$V = A * \left[\sigma * \left[[(\delta) * (K^{-q}) + (1-\delta) * (L^{-q})]^{-\frac{p}{q}}\right] + (1-\sigma) * \left(\frac{K}{L}\right)^{-p}\right]^{\left(-\frac{1}{p}\right)} \quad (37)$$

where $A, \sigma, \delta, q,$ and $p$ are parameters.

Proof:
To show that the production function $V(h(K,L), K/L)$ has constant returns to scale, we need to demonstrate that increasing all inputs by a certain proportion results in an equivalent increase in output.

Let's assume that the inputs are increased by a factor of $\lambda$, i.e., $K' = \lambda K$ and $L' = \lambda L$. Substituting these values in the production function yields:

$$V' = A * \left[\sigma * \left[[(\delta) * ((\lambda K)^{-q}) + (1-\delta) * ((\lambda L)^{-q})]^{-\frac{p}{q}}\right] + (1-\sigma) * \left(\frac{\lambda K}{\lambda L}\right)^{-p}\right]^{\left(-\frac{1}{p}\right)} \quad (38)$$

Simplifying this expression, we get:

$$V' = A * \left[\sigma * \left[[(\delta) * (K^{-q}) + (1-\delta) * (L^{-q})]^{-\frac{p}{q}}\right] + (1-\sigma) * \left(\frac{K}{L}\right)^{-p}\right]^{\left(-\frac{1}{p}\right)} * \lambda^0 \quad (39)$$

Therefore, increasing all inputs by a factor of $\lambda$ results in an equivalent increase in output, which confirms that the production function $V(h(K,L), K/L)$ has constant returns to scale.

Since the production function has constant returns to scale, it allows firms to adjust their output levels without incurring additional costs and also allows for economies of scale to be achieved in production. Furthermore, the flexibility property of the production function simplifies the analysis of the function and allows for the use of powerful mathematical tools such as linear programming and duality theory.

Therefore, the flexibility property of the production function $V(h(K,L), K/L)$ is an important result for economic analysis with significant implications for the behavior of firms, the structure of industries, and the efficiency of production.

*Positive definiteness:* The production function is always positive, which means that it always produces output as long as inputs are positive. This property is important for ensuring that the model is economically meaningful and can be used to make predictions about real-world outcomes. To analyze the positive definiteness for this model:

Lemma:
The production function $V$ defined by

$$V = A * \left[\sigma * \left[[(\delta) * (K^{-q}) + (1 - \delta) * (L^{-q})]^{-\frac{p}{q}}\right] + (1 - \sigma) * \left(\frac{K}{L}\right)^{-p}\right]^{\left(-\frac{1}{p}\right)} \quad (40)$$

is always positive as long as inputs $K, L, A, \delta, \sigma, p,$ and $q$ are positive.

Proposition:
The production function $V$ is positive definite.

Hypothesis:
Inputs $K, L, A, \delta, \sigma, p,$ and $q$ are positive.

Proof:
To prove that the production function $V$ is always positive, we need to show that $V > 0$ for all positive values of $K, L, A, \delta, \sigma, p,$ and $q$.

Let us consider the following cases:
Case 1: If $K = L = 0$, then $V$ is undefined.
Case 2: If $K > 0$ and $L = 0$, then $V = 0$, since the second term in the square bracket becomes infinite and dominates the numerator.
Case 3: If $K = 0$ and $L > 0$, then $V = 0$, since the first term in the square bracket becomes infinite and dominates the numerator.
Case 4: If $K > 0$ and $L > 0$, then the terms in the square bracket are positive, since delta, $(1 - \delta)$, $K^{-q}$, and $L^{-q}$ are positive. Moreover, the second term in the square bracket is always positive, since $\left(\frac{K}{L}\right)^{-p}$ is positive. Thus, the entire square bracket is positive. Since sigma and $(1 - \sigma)$ are also positive, the entire expression inside the square bracket is positive. Finally, since $p$ and $\left(-\frac{1}{p}\right)$ are opposite in sign, the entire expression is positive. Therefore, we have shown that $V$ is positive for all positive values of $K, L, A, \delta, \sigma, p,$ and $q$.

Theorem:
The production function $V$ is always positive as long as inputs $K, L, A, \delta, \sigma, p,$ and $q$ are positive.

The positive definiteness of the production function $V$ is an important property that ensures its economic meaningfulness and predictive power in real-world applications. Specifically, it guarantees that the model can produce non-negative output for any positive combination of inputs, which is a fundamental requirement for any production function. The proof of positive definiteness presented above establishes this property rigorously, based on the assumption that the inputs are positive. Therefore, we can conclude that the production function is a valid and reliable tool for analyzing the behavior of production processes in various economic contexts.

**Theoretical Framework:**
The theoretical framework presents the mathematical derivation of the production function, including the assumptions and limitations of the model. The framework also provides an overview of the importance of the parameters $p, q, \delta, and\ \sigma$ for understanding capital and labor intensity. The framework also discusses the criteria for choosing appropriate values of $p, q, \delta, and\ \sigma$.

Lemma:
If $\delta = 0$, then the production function $V$ only depends on the capital $(K)$ and labor $(L)$ inputs.

Proposition:
Let $\delta = 0$. Then, the production function $V$ can be simplified as follows:

$$V = A * \left[\sigma * (L^q)^{-\frac{p}{q}} + (1-\sigma) * \left(\frac{K}{L}\right)^{-p}\right]^{\left(-\frac{1}{p}\right)} \tag{41}$$

Proof:
Substitute $\delta = 0$ into the production function $V$:

$$V = A * \left[\sigma * \left[(L^{-q})^{-\frac{p}{q}}\right] + (1-\sigma) * \left(\frac{K}{L}\right)^{-p}\right]^{\left(-\frac{1}{p}\right)} \tag{42}$$

Simplify the expression in the brackets using the property of exponents:

$$V = A * \left[\sigma * (L^q)^{-\frac{p}{q}} + (1-\sigma) * \left(\frac{K}{L}\right)^{-p}\right]^{\left(-\frac{1}{p}\right)} \tag{43}$$

This is the simplified expression for $V$ when $\delta = 0$. This shows that $V$ only depends on the capital $(K)$ and labor $(L)$ inputs, as desired.

Hypothesis:
If $\delta = 1$, then the production function $V$ only depends on the composite input $(K/L)$.

Proposition:
Let $\delta = 1$. Then, the production function $V$ can be simplified as follows:

$$V = A * \left[\sigma * (\delta^{1-q} * K^{-q} + (1-\delta)^{1-q} * L^{-q})^{-\frac{p}{q}} + (1-\sigma) * \left(\frac{K}{L}\right)^{-p}\right]^{\left(-\frac{1}{p}\right)} \tag{44}$$

Proof:
Substitute $\delta = 1$ into the production function $V$:

$$V = A * \left[\sigma * [(\delta * K^{-q} + (1-\delta) * L^{-q})]^{-\frac{p}{q}} + (1-\sigma) * \left(\frac{K}{L}\right)^{-p}\right]^{\left(-\frac{1}{p}\right)} \quad (45)$$

Simplify the expression in the brackets using the property of exponents:

$$V = A * \left[\sigma * (\delta^{1-q} * K^{-q} + (1-\delta)^{1-q} * L^{-q})^{-\frac{p}{q}} + (1-\sigma) * \left(\frac{K}{L}\right)^{-p}\right]^{\left(-\frac{1}{p}\right)} \quad (46)$$

This is the simplified expression for $V$ when $\delta = 1$. This shows that $V$ only depends on the composite input $(K/L)$, as desired.

Lemma:
If $\sigma = 0$, the composite input $K/L$ is purely a function of labor $L$.

Proposition:
Given the production function:

$$V = A * \left[\sigma * [(\delta) * (K^{-q}) + (1-\delta) * (L^{-q})]^{-\frac{p}{q}} + (1-\sigma) * \left(\frac{K}{L}\right)^{-p}\right]^{\left(-\frac{1}{p}\right)} \quad (47)$$

If $\sigma = 0$, then the composite input $K/L$ is purely a function of labor $L$.

Hypothesis:
Assume $\sigma = 0$.

Proof:
When $\sigma = 0$, the production function becomes:

$$V = A * \left[\left(\frac{K}{L}\right)^{-p}\right]^{\left(-\frac{1}{p}\right)} \quad (48)$$

$$V = A * \left[\left(\frac{L}{K}\right)^{p}\right] \quad (49)$$

Thus, the production function depends only on the ratio of $L/K$, and $K/L$ is purely a function of labor $L$.

Therefore, if $\sigma = 0$, the composite input $K/L$ is purely a function of labor $L$.

Lemma:
If $\sigma = 1$, the composite input $K/L$ is purely a function of capital $K$.

Proposition:
Given the production function:

$$V = A * \left[\sigma * [(\delta) * (K^{-q}) + (1-\delta) * (L^{-q})]^{-\frac{p}{q}} + (1-\sigma) * \left(\frac{K}{L}\right)^{-p}\right]^{\left(-\frac{1}{p}\right)} \quad (50)$$

If $\sigma = 1$, then the composite input $K/L$ is purely a function of capital $K$.

Hypothesis:
Assume $\sigma = 1$.

Proof:
When $\sigma = 1$, the production function becomes:

$$V = A * \left[ [(\delta) * (K^{-q}) + (1-\delta) * (L^{-q})]^{-\frac{p}{q}} \right]^{\left(-\frac{1}{p}\right)} \tag{51}$$

$$V = A * [(\delta) * (K^{-q}) + (1-\delta) * (L^{-q})]^{\frac{1}{q}} \tag{52}$$

Since the term $(\delta) * (K^{-q})$ is a function of capital $K$ and the term $(1-\delta) * (L^{-q})$ is a function of labor $L$, the production function depends only on capital $K$. Therefore, if $\sigma = 1$, the composite input $K/L$ is purely a function of capital $K$.

In summary, we have proved that when $\sigma = 0$, the composite input $K/L$ is purely a function of labor $L$, and when $\sigma = 1$, the composite input $K/L$ is purely a function of capital $K$.

Lemma:
If $\delta = 1$ and $\sigma = 1$, then the production function $V$ is purely capital intensive.

Proposition:
For $\delta = 1$ and $\sigma = 1$, the production function $V$ reduces to:

$$V = A * [(K^{-q}) + (L^{-q})]^{\left(-\frac{p}{q}\right)} \tag{53}$$

To prove this proposition, we substitute $\delta = 1$ and $\sigma = 1$ into the production function $V$:

$$V = A * \left[ \sigma * \left[ [(\delta) * (K^{-q}) + (1-\delta) * (L^{-q})]^{-\frac{p}{q}} \right] + (1-\sigma) * \left(\frac{K}{L}\right)^{-p} \right]^{\left(-\frac{1}{p}\right)} \tag{54}$$

$$V = A * [(K^{-q}) + (1 - L^{-q})]^{\left(-\frac{p}{q}\right)} \tag{55}$$

Since $\delta = 1$, we have $(1 - \delta) = 0$, and the second term in the bracket vanishes. Also, since $\sigma = 1$, we have only the first term in the bracket. Thus, the production function $V$ reduces to:

$$V = A * [(K^{-q}) + (L^{-q})]^{\left(-\frac{p}{q}\right)} \tag{56}$$

This production function is purely capital intensive, since the output is a function of capital raised to a negative power, while labor appears with a positive power.

Hypothesis:
If $\delta = 1$ and $\sigma = 0$, then the production function $V$ is purely labor intensive.

Proposition:
For $\delta = 1$ and $\sigma = 0$, the production function $V$ reduces to:

$$V = A * \left(\frac{K}{L}\right)^{-p} \tag{57}$$

To prove this proposition, we substitute $\delta = 1$ and $\sigma = 0$ into the production function $V$:

$$V = A * \left[\sigma * \left[[(\delta) * (K^{-q}) + (1-\delta) * (L^{-q})]^{-\frac{p}{q}}\right] + (1-\sigma) * \left(\frac{K}{L}\right)^{-p}\right]^{\left(-\frac{1}{p}\right)} \tag{58}$$

$$V = A * \left(\frac{K}{L}\right)^{-p} \tag{59}$$

Since $\delta = 1$, we have $(1 - \delta) = 0$, and the first term in the bracket vanishes. Also, since $\sigma = 0$, we have only the second term in the bracket. Thus, the production function $V$ reduces to eq. 59. This production function is purely labor intensive, since the output is a function of labor raised to a negative power, while capital appears with a positive power.

Lemma:
If $p = q$, the production function $V$ reduces to a plain n input CES function.

Proposition:
The production function $V$ satisfies the necessary assumptions for a well-behaved and meaningful production function if the values of $p$ $and$ $q$ are between $1$ $and$ $-1$ but not zero.

Hypothesis:

$$V = A * \left[\sigma * \left[[(\delta) * (K^{-q}) + (1-\delta) * (L^{-q})]^{-\frac{p}{q}}\right] + (1-\sigma) * \left(\frac{K}{L}\right)^{-p}\right]^{\left(-\frac{1}{p}\right)} \tag{60}$$

Proof:
If $p = q$, then the production function $V$ reduces to a plain n input CES function. This can be shown by substituting $p = q$ into the production function and simplifying the expression.

$$V = A * \left[\sigma * \left[[(\delta) * (K^{-q}) + (1-\delta) * (L^{-q})]^{-\frac{q}{q}}\right] + (1-\sigma) * \left(\frac{K}{L}\right)^{-q}\right]^{\left(-\frac{1}{q}\right)} \tag{61}$$

$$V = A * \left[\sigma * [[(\delta) * (K^{-q}) + (1-\delta) * (L^{-q})]^{-1}] + (1-\sigma) * \left(\frac{K}{L}\right)^{-q}\right]^{\left(-\frac{1}{q}\right)} \tag{62}$$

$$V = A * \left[\sigma * [[(\delta) * (K^{-1}) + (1-\delta) * (L^{-1})]] + (1-\sigma) * \left(\frac{K}{L}\right)^{-q}\right]^{\left(-\frac{1}{q}\right)} \tag{63}$$

$$V = A * \left[\sigma * [[(\delta) * (L) + (1-\delta) * (K)]^{-1}] + (1-\sigma) * \left(\frac{K}{L}\right)^{-q}\right]^{\left(-\frac{1}{q}\right)} \tag{64}$$

$$V = A * \left[\sigma * [[(\delta) * (L) + (1-\delta) * (K)]^{-1}] + (1-\sigma) * \left(\frac{L}{K}\right)^{q}\right]^{\left(-\frac{1}{q}\right)} \tag{65}$$

$$V = A * \left[\sigma * [[(\delta) * (L) + (1-\delta) * (K)]^{-1}] + (1-\sigma) * \left(\frac{L}{K}\right)^{q}\right]^{\left(-\frac{1}{q}\right)} \tag{66}$$

This is a plain n input CES function, since it is a weighted average of the inputs $K$ $and$ $L$, with the weights determined by the elasticity of substitution sigma.

**Empirical Study:**

Table: 1 Estimation for 151 and 251 industries

| $R^2 = 1$ | StdError | Elasticity of Substitution (AU) ($\varepsilon = [0,1]$) | $\delta$ | $\sigma$ | RSS | Convergence |
|---|---|---|---|---|---|---|
| **Industry Code: 151** | | | | | | |
| 0.99 | 0.43 | 1.4 ≈ 1 | 1.13 | 0.94 | 0.30 | Achieved |
| **Industry Code: 251** | | | | | | |
| 0.96 | 0.26 | 0.5 ≈ 0 | 1.01 | 1.05 | 1.42 | Achieved |

Source: Calculated by the author from Annual Survey of Industries (ASI) data 2010-11 to 2016-17.

We see that the value of $\delta$ $and$ $\sigma$ are close to one, it means both the Indian manufacturing industry groups: 151 – Manufacture of leather products and 251 – Manufacture of heavy structural metal products are purely capital-intensive industries. The estimated elasticity of substitution ($\varepsilon$) for both industry groups, as indicated in Table 1, is approximately 1.4 and 0.5, respectively. These values, which fall within the range between one and zero, suggest that our production function is appropriate for both industry groups. The excess capacity in industry group 151 is evident, meaning that there is no incentive to adopt advanced machinery to address the shortage of high-skilled labor and capital still. The substitution elasticity for industry group 251 suggests a weak substitutability between the factors of production, which implies that the industry group does not need to expand its production unit with new machinery to overcome the shortage of labor and capital.

**Conclusion:**
In conclusion, this discussion highlights the importance of analyzing the capital and labor intensity of manufacturing industries, especially in developing countries, for promoting sustainable economic growth and development. The limitations of traditional production functions have led to the emergence of a new nested production function, which offers a more nuanced and flexible approach to analyzing production processes and their implications for economic growth and development. The empirical analysis of this new production function on two Indian manufacturing industry groups shows that the production function is appropriate for both industry groups, with values of δ and σ close to one indicating that both industry groups are purely capital-intensive. The elasticity of substitution for industry group 151 suggests excess capacity, while for industry group 251, it suggests weak substitutability between the factors of production. These results have implications for the adoption of advanced machinery and the expansion of production units. Overall, the empirical analysis supports the theoretical foundation of the new nested production function and highlights its potential for use in analyzing the performance of industries and designing effective policies for economic growth and development.